\documentclass[aps,prb,twocolumn,superscriptaddress,floatfix]{revtex4}
\usepackage{graphicx}
\usepackage{psfrag}
\usepackage{color}
\usepackage{subfigure}
\usepackage{amsmath}
\definecolor{dred}{rgb}{0.7,0.0,0.0}
\allowdisplaybreaks 
 
\begin{document}

\title{Study of the  magnetic state of K$_{0.8}$Fe$_{1.6}$Se$_2$ \\
using the five-orbital Hubbard model in the Hartree-Fock approximation   }

\author{Qinlong Luo}
\author{Andrew Nicholson}
 
\affiliation{Department of Physics and Astronomy, The University of
  Tennessee, Knoxville, TN 37996} 
\affiliation{Materials Science and Technology Division, Oak Ridge
  National Laboratory, Oak Ridge, TN 32831} 

\author{Jos\'e Riera}
\affiliation{Instituto de F\'isica Rosario, Universidad Nacional de Rosario, 2000-Rosario, Argentina}

\author{Dao-Xin Yao}
\affiliation{State Key Laboratory of Optoelectronic Materials
and Technologies, Sun Yat-sen University, Guangzhou 510275, China}

\author{Adriana Moreo}
\author{Elbio Dagotto}

\affiliation{Department of Physics and Astronomy, The University of
  Tennessee, Knoxville, TN 37996} 
\affiliation{Materials Science and Technology Division, Oak Ridge
  National Laboratory, Oak Ridge, TN 32831}

\date{\today}

\begin{abstract}

Motivated by the recent discovery of Fe-based superconductors 
close  to an antiferromagnetic insulator in the
experimental phase diagram, here the five-orbital
Hubbard model (without lattice distortions) 
is studied using the real-space Hartree-Fock approximation,
employing a 10$\times$10 Fe cluster with Fe vacancies 
in a $\sqrt{5}\times\sqrt{5}$ pattern. 
Varying the Hubbard and Hund couplings, and at electronic density $n$=6.0, 
the phase diagram contains 
an insulating state with the same spin pattern as observed experimentally, involving
2$\times$2 ferromagnetic plaquettes coupled with one another 
antiferromagnetically. The presence of local FM tendencies 
is in qualitative agreement with
Lanczos results for the three-orbital model also reported here. 
The magnetic moment $\sim$3$\mu_B$/Fe
is in good agreement with experiments. 
Several other phases are also stabilized in the phase
diagram, in agreement with recent calculations using 
phenomenological models.


\end{abstract}

\maketitle


{\it Introduction.} 
Among the most
recent exciting developments in the field of Fe-based superconductors~\cite{johnston} 
is the discovery of superconductivity (SC) with $T_c$$\sim$30~K in the heavily
electron-doped 122 iron-chalcogenides K$_{0.8}$Fe$_{2-x}$Se$_2$ 
and (Tl,K)Fe$_{2-x}$Se$_2$ compounds.\cite{guo}
These materials contain ordered Fe vacancies in the FeSe layers, 
increasing the complexity of these systems.
Recent neutron scattering results 
for the parent compound K$_{0.8}$Fe$_{1.6}$Se$_2$,\cite{bao} with
the Fe-vacancies arranged in a $\sqrt{5}\times\sqrt{5}$ pattern,
revealed an unexpected  magnetic and insulating state 
involving 2$\times$2 Fe plaquettes that have their four Fe spins
ferromagnetically ordered, and with these plaquettes 
coupled to each other antiferromagnetically.\cite{ricci}
The ordered magnetic moment is 3.31~$\mu_B$/Fe, the largest
among all Fe pnictide and chalcogenide superconductors, and the magnetic
transition occurs at a high temperature 
$T_N$$\approx$559~K. 
Angle-resolved photoemission experiments for (Tl, K)Fe$_{1.78}$Se$_2$
have revealed a Fermi surface (FS) with only
electron-like pockets at the $(\pi,0)$ and $(0,\pi$) points and a
nodeless superconducting gap at those pockets.\cite{ding} 
The superconducting phase in these compounds cannot be explained by  
the nesting between hole and electron pockets.\cite{ding,petrovic}
Moreover, the resistivity of these materials displays a behavior 
corresponding to an insulator
in a robust range of the Fe concentration $x$,\cite{fang}  
suggesting that SC may arise
from the doping of a Mott insulator, as in the cuprates. All these results 
certainly have challenged 
prevailing ideas for the origin of SC in these
materials that were originally based on a nested FS 
picture and a metallic parent state.

Several theoretical efforts have recently addressed the exotic magnetic state
that appears in the presence of vacancies. 
Band structure calculations described this
state as an antiferromagnetic insulator
with a gap $\sim$0.4-0.6~eV.\cite{lda,lda2} 
Several model Hamiltonian calculations have also been presented and,
in particular, two recent publications are important
to compare our results against. 
Yu {\it et al.}\cite{rong+si} analyzed this problem
using a phenomenological $J_1$-$J_2$ spin 
model (see also Ref.~\onlinecite{lda}) 
with nearest-neighbors (NN) and next-NN terms superexchange couplings,
studied via classical Monte Carlo. In this analysis the couplings inside
the 2$\times$2 plaquettes and those between plaquettes were allowed to be 
different, and also
to take positive or negative values.
Five antiferromagnetic phases, 
including the phase found experimentally\cite{bao} in 
K$_{0.8}$Fe$_{1.6}$Se$_2$,
which was dubbed ``AF1'', were found varying the $J_1$
  and $J_2$ couplings.\cite{ju} 
From a 
different perspective that relies  on a two-orbital ($d_{xz}$ and $d_{yz}$) 
spin-fermion model for pnictides, 
and with tetramer lattice distortion incorporated, 
Yin {\it et al.}\cite{BNL} 
studied the regime of electronic density $n$=1 (one electron per Fe), where they also reported 
the presence of an AF1 state, found competing with a ``C'' type state
with wavector $(\pi,0)$. 

\begin{figure}[thbp]
\begin{center}
\includegraphics[width=9.0cm,clip,angle=0]{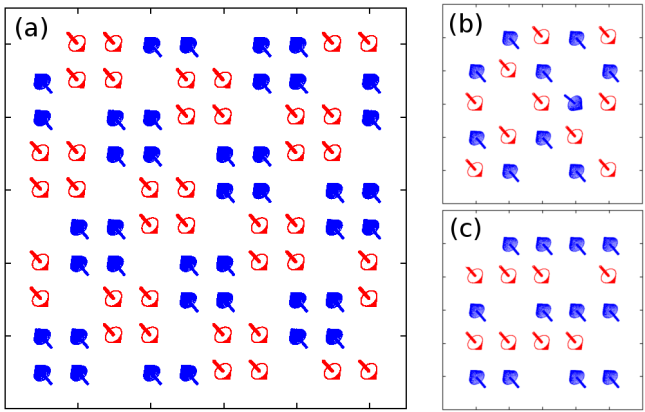}
\caption{(Color online) 
(a) Sketch of the AF1 state found to be stable in a
region of the $U$-$J/U$ phase diagram (see Fig.~2) in our
HF approximation to the five-orbital Hubbard model,
in agreement with neutron diffraction.\cite{bao} (b) A competing
state dubbed AF4 (stable at smaller $J/U$'s in Fig.~2). 
(c) The C competing state.
For (b) and (c), a subset of the 10$\times$10 
cluster used is shown.
}
\vskip -0.5cm
\label{fig-spins}
\end{center}
\end{figure}

In the present publication, a more fundamental five-orbital Hubbard model, without
lattice distortions, is
investigated. Our main result is that increasing the Hubbard coupling $U$ and
the Hund coupling $J$, a robust region of stability of the AF1 state is found.
Our effort allows to display the regions of dominance 
of the many competing states in terms of $U$ and $J/U$,
facilitating a discussion on possible phase transitions among these states
by varying experimental parameters. 
A sketch of the AF1 state and its two main competitors,
the C and AF4 states, is in Fig.~\ref{fig-spins}.
Our results agree qualitatively in several respects 
with the phenomenological studies of Refs.~\onlinecite{rong+si,BNL} particularly
if a combination of results of these 
investigations is made. Finally, also note that a recent study~\cite{illinois} of the
three-orbital Hubbard model~\cite{three} using 
mean-field techniques~\cite{luo} has also reported the existence 
of an AF1 state but with orbital order (OO). 
The relation with our results will also be discussed below.

{\it Models and methods.} In this manuscript, 
the standard multiorbital Hubbard model will be used. This model
has been extensively described in several
previous publications, by our group and others.
More specifically, the model used is the 
five-orbital Hubbard model defined explicitly in Ref.~\onlinecite{luo} 
with the hopping amplitudes introduced by Graser {\it et al.}~\cite{graser} 
By construction, this model 
has a FS that is in close agreement with band structure
calculations and angle-resolved photoemission results for the pnictides
without vacancies. The presence of the realistic AF1 state in our results,
as shown below, suggests that the same set of hopping amplitudes can
be used in a system with Fe vacancies. The electronic density 
will be $n$=6.0, i.e. 6 electrons 
per Fe, for all the five-orbital model results presented below. The
couplings are the on-site Hubbard repulsion $U$ at the same orbital and
the on-site Hund coupling $J$. The on-site inter-orbital repulsion $U'$ 
satisfies $U'$=$U$-$2J$.
The computational method that is employed to extract information from this five-orbital
model relies on the study of a 10$\times$10 cluster, 
as sketched in Fig.~\ref{fig-spins}(a), using
periodic boundary conditions. In this cluster, several vacancies
and 2$\times$2 building blocks fit comfortably inside, 
giving us confidence that the main
local tendencies to magnetic order are not dramatically affected by size effects. 

With regards to the actual many-body technique
used to study the 10$\times$10 cluster, 
here the real-space Hartree-Fock (HF) approximation was employed. The method
is a straightforward generalization of that used recently 
by our group in Ref.~\onlinecite{luo-stripes} in
the study of charge stripe tendencies for the two-orbital model.
This HF real-space approach was preferred over a momentum-space procedure in order 
to allow for the system to select spontaneously
the state that minimizes the HF energy, at least 
for the finite cluster here employed. In practice, the many fermionic
expectation values that appear in the HF formalism must be found iteratively
by energy minimization.
At the beginning of the iterative process, both random initial conditions
as well as initial ordered states favoring 
the many phases that are anticipated to be in competition were employed.
After each of the computer runs using different initial conditions have 
reached convergence, at a fixed $U$ and $J/U$, a mere
comparison of energies allowed us to find the ground state for those 
particular couplings. In our setup, typical
running times for one set of couplings $U$-$J/U$ required 
approximately 20 hours of CPU time to reach convergence.\cite{numerical} 
Dozens of computer cluster nodes have been used to complete
our analysis in a parallel manner.

\begin{figure}[thbp]
\begin{center}
\includegraphics[width=10.0cm,clip,angle=0]{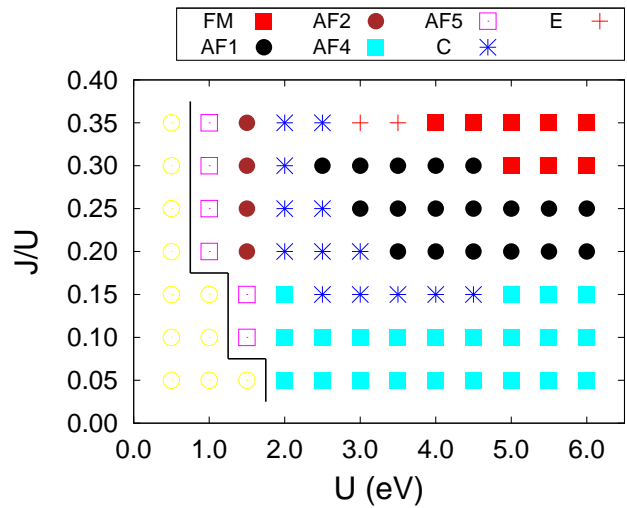}
\vskip -0.3cm
\caption{(Color online) 
Phase diagram of the five-orbital Hubbard model with $\sqrt{5}$$\times$$\sqrt{5}$ 
Fe vacancies studied via the real-space HF
approximation to a 10$\times$10 cluster, employing the procedure for convergence described
in the text. With increasing $U$, clear
tendencies toward magnetic states are developed. The realistic
AF1 state found in neutron scattering experiments\cite{bao} appears here above
$J/U$=0.15 and for $U$ larger than 2.5~eV. The notation 
for the most important states
is explained in Fig.~\ref{fig-spins} and 
for the rest in Refs.~\onlinecite{lda,rong+si,BNL}. The region with low-intensity yellow circles
at small $U$ is non-magnetic.\cite{comment} 
}
\vskip -0.4cm
\label{fig-phase.diagram}
\end{center}
\end{figure}

{\it Results.}
The main results arising from the computational minimization process just described
are summarized in the phase diagram shown in Fig.~\ref{fig-phase.diagram}.
Since the hopping parameters of Ref.~\onlinecite{graser} are already in eV units,
our Hubbard coupling $U$ is also displayed in the same units. The notation for the
many competing phases used here is
that of Refs.~\onlinecite{lda,rong+si,BNL} to facilitate comparisons.
The main result of the present work 
is that  our phase diagram displays a robust region where 
the magnetic order unveiled by neutron diffraction,\cite{bao} 
see Fig.~\ref{fig-spins}(a), is found to be stable.
The ratio $J/U$ needed for the AF1 phase to be the ground state is in good
agreement with previous estimations for the same model, although obtained in the
absence of vacancies, based on the comparison of
Hubbard model results against neutron and photoemission data.\cite{luo}
The ratio $J/U$ is surprisingly similar between 
the pnictides and the chalcogenides.
With regards to the actual value of $U$ in eV's, 
the range unveiled in previous investigations that
focused on the ``1111'' and ``122'' families of pnictides was 
approximately 1.5 ~eV (see Fig.~13 of Ref.~\onlinecite{luo}). The increase to
2.5~eV in the present investigation is not surprising in view of 
the more insulating characteristics of materials 
such as K$_{0.8}$Fe$_{1.6}$Se$_2$, and suggests that merely adding vacancies to
the intermediate $U$ state of the pnictides (without vacancies) 
is not sufficient to stabilize
the AF1 state but an increase in $U$ is also needed. 
Finally, with regards to OO,
none is observed in the AF1 state in the range of $U$ shown in Fig.~\ref{fig-phase.diagram}, 
i.e. for $U$$\leq$6~eV. In this range,
the electronic density of all the orbitals ($d_{xz}$ and $d_{yz}$ in particular) is independent
of the site location in the cluster analyzed.
However, upon further
increasing $U$ to 8~eV and beyond, the {\it same} OO pattern found
in the three-orbital model~\cite{illinois} appears 
in our calculations (not shown explicitly), with the populations of the
 $d_{xz}$ and $d_{yz}$ orbitals now being different at all sites. It seems
that with five orbitals the AF1 state manifests itself both with and without
OO, depending on $U$, while for three orbitals the intermediate phase
with AF1 magnetic order and without OO is not present.\cite{illinois}

\begin{figure}[thbp]
\begin{center}
\vskip -0.5cm
\includegraphics[width=9.0cm,clip,angle=0]{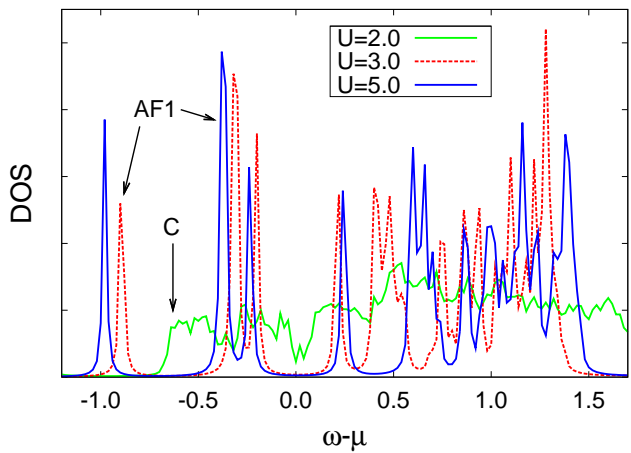}
\caption{(Color online) Density of states of the AF1 and C phases
sketched in Figs.~\ref{fig-spins}(a,c), at the $U$'s indicated, 
$J/U$=0.25, and using a 10$\times$10 cluster. 
The gap at the chemical potential suggests that the AF1 state ($U$=3 and 5) 
is an insulator, although with a mild $U$ dependence in the value of this gap.
On the other hand, the C state appears to have only a pseudogap at the Fermi
level.\cite{comment2}
}
\vskip -0.6cm
\label{fig-dos}
\end{center}
\end{figure}

Together with the realistic AF1 phase, Fig.~\ref{fig-phase.diagram} reveals several other states, and two
of them are prominent. Keeping the ratio $J/U$ constant but reducing $U$, the previously described 
C-type state (Fig.~\ref{fig-spins}(c)) was found to be stable. 
This is reasonable since without Fe vacancies this state is
the dominant spin order in the intermediate range of couplings, where 
the ground state is both
metallic and magnetic.\cite{luo} 
In K$_{0.8}$Fe$_{1.6}$Se$_2$, as the bandwidth is increased by, e.g.,
increasing the pressure, a transition from the AF1 to the C-state 
could be experimentally observed. In these
regards, our conclusions agree with Ref.~\onlinecite{BNL} that 
the C-state is the main competitor of the AF1 state. 
However, note that other states reported 
in Ref.~\onlinecite{rong+si} are also present in our phase
diagram. For instance, the AF4 state (Fig.~\ref{fig-phase.diagram}(b)) 
is stable in a large region of parameter space at small values of $J/U$.
Thus, overall our results support a combination of
the main conclusions of Refs.~\onlinecite{rong+si,BNL}.

The density-of-state (DOS) for the AF1 phase 
is shown in Fig.~\ref{fig-dos} for representative couplings. The presence of a gap
at the chemical potential indicates an insulating state, in
agreement with experiments.\cite{bao} 
This is not surprising considering
that the transport of charge from each 2$\times$2 building block to a
NN block may be suppressed due to the effective 
antiferromagnetic coupling between blocks, at least at large $U$ and $J$. 
In other words, using a tilted square lattice made out 
of 2$\times$2 superspin blocks, the state is actually a staggered
antiferromagnet that is known to have low conductance. On the other hand, it is
interesting to observe that the AF1 gap is only weakly dependent on $U$, 
suggesting that not only the increase in $U$ is responsible
for the insulating behavior but there must be other
geometrical reasons that may contribute to the gap through quantum interference.
This is reminiscent of results reported years ago for the insulating CE phase
of half-doped manganites, state that is stabilized in the phase diagram even in the
absence of electron-phonon coupling due to the peculiar geometry of the
zigzag chains involved in the CE state and the multi-orbital nature of the problem, 
that induces a band insulating behavior.\cite{hotta}
Thus, in agreement with recent independent observations,\cite{BNL} our results suggest
that the insulator stabilized in the presence of Fe vacancies may have a dual
Mott and band-insulating character. Note also that the competing C-state only has
a pseudogap (Fig.~\ref{fig-dos}), and thus it may be a bad metal.\cite{comment2}

With regards to the strength of the FM tendencies in each
of the 2$\times$2 building blocks of the AF1 state, examples of the
values of the magnetic moment $m$ (in Bohr magnetons, assuming $g$=2, and at $J/U$=0.25) 
are $m$=3.87 ($U$=3.0),
$m$=3.93 ($U$=4.0), and $m$=3.95 ($U$=5.0), in good
agreement with neutron diffraction results\cite{bao} $m$=3.3.
Thus, the Fe spins in the AF1 superblocks are near the saturation value 4.0~$\mu_B$ at $n$=6.0.
Note that the competing C-phase also has a surprisingly large moment $m$=3.5 at $U$=2.0 and $J/U$=0.25.

\begin{figure}[thbp]
\begin{center}
\includegraphics[width=8.5cm,clip,angle=0]{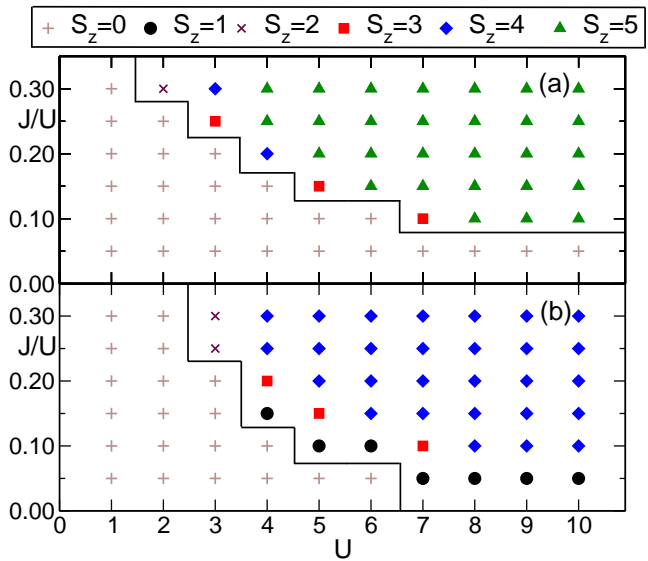}
\caption{(Color online) 
Total spin 
of a 2$\times$2 cluster using the three-orbital Hubbard model\cite{three} and
the Lanczos technique, varying $U$ and $J/U$. Panel (a) is for 10 electrons,
while (b) is for 14 electrons. The ``undoped'' limit 
is for 16 electrons in the three-orbital model.\cite{three} 
The figure shows that in these hole doped clusters FM tendencies 
develop as $U$ grows, at realistic $J/U$'s, compatible 
with the five-orbital HF results.
}
\vskip -0.6cm
\label{fig-andrew}
\end{center}
\end{figure}

{\it Results for the three-orbital Hubbard model.}
The results reported thus far 
have been obtained 
under the HF approximation. Better unbiased approximations for this model 
are not currently available. However, at least
consistency checks of the present results can be carried out 
using the Lanczos
technique restricted to the 2$\times$2 cluster of irons 
that forms the AF1 state. For our problem, 
an additional simplification from five to three orbitals ($d_{xz}$, $d_{yz}$,
and $d_{xy}$) is needed 
to reduce the Hilbert space to a reasonable size, thus here the model
introduced by Daghofer {\it et al.}\cite{three} was used.
The present Lanczos study
is equivalent to a 12-sites one-orbital Hubbard model 
which can be done comfortably with present day computers even with the
open boundary conditions (OBC) employed here.\cite{andrew} 
Our focus has 
been on the total spin quantum number to search for indications of
FM tendencies in the 2$\times$2 cluster. 
The main results are in Fig.~\ref{fig-andrew} 
for the case of hole doping.
These results indicate that with increasing $U$ and $J/U$,
FM tendencies indeed develop, in agreement 
with the five-orbital HF results (Fig.~\ref{fig-phase.diagram}). 
With electron doping also FM tendencies were found (not shown). 
In these Lanczos results the transition from low- to high-spin 
is rather abrupt. However, note that in the 2$\times$2 cluster
there is obviously no room to distinguish a fully FM state from an AF1 state, 
thus the large-spin region of the 2$\times$2 cluster may correspond 
to any of the two if
larger clusters could be studied. Nevertheless, 
solving exactly this case
allows us to confirm that 
with increasing $U$ (at realistic $J/U$'s), there
is a clear tendency in the multiorbital Hubbard model 
towards local FM order
in 2$\times$2 clusters.

{\it Summary.} Real-space HF-approximation results for the five-orbital 
Hubbard model, supplemented by Lanczos calculations for three-orbital on
a 2$\times$2 plaquette, have been
presented for the case of a $\sqrt{5}\times\sqrt{5}$ arrangement of Fe vacancies. The phase
diagram obtained by varying $U$ and $J/U$ 
contains the magnetic state found in 
neutron diffraction experiments.\cite{bao} 
 This state arises at intermediate couplings $U$ and $J/U$, and
in the phase diagram it is not in 
contact with the paramagnetic metallic state of the weak
coupling limit. Thus, FS nesting cannot explain the stability
of the AF1 magnetic state in the presence of Fe vacancies.\cite{BNL}
The density of states shows that the AF1 state is an insulator, 
but since the gap does not present a strong dependence on $U$ its origin may
reside in a combination of Hubbard and band-insulator features. 
In agreement with recent spin\cite{rong+si} and spin-fermion\cite{BNL} 
model calculations, several other 
magnetic phases were found here,
suggesting that transitions among these
competing states, or among AF1 with and without OO,\cite{illinois} 
could be observed experimentally particularly by
modifications in the carrier's bandwidth.

{\it Acknowledgments.} Work supported by 
the U.S. Department of Energy, Office of Basic Energy Sciences,
Materials Sciences and Engineering Division (Q.L., A.N., A.M., E.D.),
CONICET, Argentina (J.R.), and the NSFC-11074310 and Fundamental
Research Funds for the Central Universities (D.X.Y.).
The computational studies used the Kraken supercomputer
of the National Institute for Computational
Sciences.


\begin{thebibliography}{10}
\bibitem{comment} However, a very small
FM moment $m$$\sim$0.02 systematically appeared in our numerical study in
portions of the ``non-magnetic'' small $U$ region. This subtle effect 
will be investigated in future efforts since it is unrelated to the large
moments found in K$_{0.8}$Fe$_{1.6}$Se$_2$.
\bibitem{comment2} The competing AF4 state was found
to be an insulator (not shown) with a large gap $\sim$1.4~eV.
\end{thebibliography}
\end{document}